# Hole Mobility and Scattering Mechanism on Undoped MOCVD Grown of GaSb

*Altje Latununuwe[1>], Pepen Arifin[2>], Euis Sustini[2>], and Horasdia Saragih[2,3>]*
*1> Department of Physics Education, FKIP Pattimura University, Ambon, Indonesia*
*2> Laboratory of Electronic Material Physics , Department of Physics ITB, Bandung, Indonesia*
*3> Department of Physics, Pattimura University, Ambon, Indonesia*
*E-mail: Altje@mailcity.com*

***Abstract***
*The temperature variation of partial mobility and effective mobility for undoped p-GaSb have been calculated using Matthiessen's rule. The effective mobility depends on the ionized, acoustic, polar optical and non-polar optical scattering processes. The analysis shows that the effective mobility for undoped p-GaSb at room temperature is about 905 $cm^2/Vs$ with the hole concentration of about $10^{16}$ $cm^{-3}$. The dependence of mobility on the hole concentration at 300 K has been analyzed and shows that mobility decreases with increasing hole concentrations. The result of this analysis then is used to analyze the electronic properties of thin film undoped p-GaSb which grown by MOCVD method. The films have been characterized using Hall-Van der Pauw method to determine the mobilities and hole concentrations. Comparison between experimental data and the theoretical analysis shows that the films have the high level of impurities. Therefore the four scattering mechanisms are governed the hole mobility of undoped p-GaSb at room temperature.*

## 1. Introduction

Gallium antimonide (GaSb) is a semiconductor material which has the highest mobility among the III-V compound semiconductor materials. Because of its high mobility, GaSb can be applied as HFET (Heterojunction Field Effect Transistor) technology such as InAs/AlSb/GaSb compound where the function of GaSb is as p-channel[1)]. Based on the matters mentioned above, the transport properties of GaSb such as the scattering processes that influence GaSb mobility have been the subject of investigation for the past three decades. Several workers such as Dutta et al investigated the transport properties of GaSb grown by using Vertical Bridgman Technique[2)]. In p-GaSb material, the most important scattering processes involve interactions of holes with lattice vibrations ( phonons ) and with impurity atoms. The scattering processes associated with impurities and phonons are acoustic, non-polar, and polar optical phonons and ionized impurity scatterings.[2)]

In this paper, we report the hole mobility and hole concentrations of undoped metalorganic chemical vapor deposition (MOCVD) grown of GaSb at room temperature that characterized using Hall-Van der Pauw method. In addition, the theoretical mobility as a function of temperature and concentration has been calculated to assess the quality of the films. Comparison with the experimental mobilities established the nature of the scattering processes governing the mobility.

## 2. Experimental Procedure

The growth of undoped p-GaSb films was carried out in a vertical-type low pressure MOCVD system. Trimethylgallium (TMGa) and tridismethylaminoantimony (TDMASb) were used as the source materials. Semi-insulating GaAs wafers were used as substrates. The growth temperatures were from 500 – 540 $^0$ C.

The electrical properties of the films at room temperature were evaluated by Hall effect measurements in the van der Pauw configuration. Prior to measurement, the ohmic contacts on these films were formed by evaporation of Ag .

## 3. Theoretical Analysis

A relaxation time approximation is widely used to analyze the scattering processes and calculate the mobility. The relaxation times for acoustic ($\tau_{AC}$), non polar ($\tau_{NPO}$), and polar ($\tau_{PO}$) optical phonons and ionized impurity ($\tau_{ION}$) scattering are given below[2)]:

$$\tau_{AX} = \left( \frac{\pi (\eta)^4 \rho \bar{u}^2}{\sqrt{2} E_{AX}^2 (m_d)^{3/2} kT} \right) E^{-1/2} \qquad (1)$$

$$\tau_{NPO} = \left( \frac{\sqrt{2}\pi(\eta)^4 \rho \bar{u}^2}{E_{NPO}^2 (\eta)\omega_0 (m_d)^{3/2}} \right)$$

$$\begin{bmatrix} N_0 (E + h\omega_0)^{1/2} \\ + \mathrm{Re}(E - h\omega_0)^{1/2} (N_0 + 1) \end{bmatrix}^{-1} \quad (2)$$

$$\tau_{PO} = \left( \frac{m_d}{2} \right)^{1/2} \frac{\chi(\theta / T) E^{1/2}}{eE_{PO} N_0} \quad (3)$$

$$\tau_{ION} = \left( \frac{(4\pi K_s)^2 (2 m_d)^{1/2}}{\pi e^4 N_I \phi(\gamma)} \right) E^{3/2} \quad (4)$$

where ρ is the density, $\bar{u}$ is the average speed of sound in the material, $\eta\omega_0$ is the optical phonon energy, and e is the electronic charge.

$N_o=[\exp(\theta/T)-1]^{-1}$ is the number of optical phonons. Here $\theta = \eta\omega_0/k$ is the optical phonon Debye temperature and k is the Boltzmann constant. $E_{AC}$ and $E_{NPO}$ are the deformation potentials for acoustic and non-polar scattering and $m_d$ is the density of states mass. In equation (3), $E_{PO}$ is defined as

$$E_{PO} = \frac{m_d e \omega_0}{2h} \left( \frac{1}{K_\infty} - \frac{1}{K_s} \right) \quad (5)$$

where $K_\infty$ and $K_s$ are the high frequency and static dielectric constants, respectively. Also $\chi(\theta/T)$ in equation (3) is defined as a function based on the approximation that is done by H. Ehrenreich.(3) In equation (4), $\phi(\gamma)$ is defined as

$$\phi(\gamma) = \ln(1 + \gamma^2) - \frac{\gamma^2}{1 + \gamma^2} \quad (6)$$

with $\gamma^2 = 8m_d E L_D^2/\eta^2$, where $L_D^2=K_s kT/e^2p^*$ and $p^* = p_1 + p_2$. $N_I$ is the number of ionized scatters and is equal to $(p_1 + p_2 + 2N_D)$ for heavy holes and $(2p_1 + p_2 + 2N_D)$ for the light holes.

The concentrations of light-hole $p_1$ and heavy-hole $p_2$ are given by:

$$p_1 = 2(2\pi m_{d1}kT/h^2)^{3/2}\exp(E_F/kT) \quad (7)$$
$$p_2 = 2(2\pi m_{d2}kT/h^2)^{3/2}\exp[(E_F - \Delta E)/kT] \quad (8)$$

where ΔE is the energy separation between the heavy- and light-hole band maxima and its value is 20 meV.[2,4,5)]

In GaSb, there are the heavy- and light-hole bands. So, the contributions from both the light-hole and heavy-hole bands to various transport properties should be taken into account.[2)]. Thus to calculate the mobility for each band is given by[2)]:

$$\mu_i = \frac{e < \tau_i(x) >}{m_{ci}} \quad (9)$$

where the subscript i = 1,2 for the two bands. $m_{ci}$ are the conductivity masses for the two bands, and $<\tau_i>$ is the relaxation time average that is given in terms of reduced the carrier energy (E/kT) as[6)]:

$$< \tau_i >= \frac{2}{3} \frac{\int_0^\infty \tau_i \left( \frac{-\partial f_0}{\partial x} \right) x^{3/2} dx}{\int_0^\infty f_0 x^{1/2} dx} \quad (10)$$

with $f_0$ is the Fermi – Dirac distribution function. In addition, $\phi(\gamma)$ (equation. 6) is a function of the energy E but, the variation of energy is slow. So, it can be moved outside the integral and evaluated at a constant energy, $E = E_m$ when computing $<\tau_i>$ (equation. 10). The value of $E_m$ is 3kT, which maximizes the integrand.[7)]

The combined mobility for the two bands is given by

$$\mu = \frac{\mu_1^2 p_1 + \mu_2^2 p_2}{\mu_1 p_1 + \mu_2 p_2} \quad (11)$$

with $\mu_1$ and $\mu_2$ are the light-and heavy-hole mobility, respectively.

The relaxation times of the individual scattering mechanisms together its reciprocal will give an effective relaxation time. Using Matthiessen's rule, the effective mobility can be calculated based on the effective relaxation time and this mobility is given by [7,8,,9,10)]:

$$\frac{1}{\mu_{eff}} = \frac{1}{\mu_{AC}} + \frac{1}{\mu_{NPO}} + \frac{1}{\mu_{PO}} + \frac{1}{\mu_{ION}} \quad (12)$$

In order to calculate the mobility, relevant material parameters are used and they are listed in the table.1.

Table.1. Material parameters of GaSb used in the calculation

| Parameters | Values | References |
|---|---|---|
| $m_d$ | $0.82m_0$ | [2,4,5] |
| $m_{c1}$ | $0.3m_0$ | [4,5] |
| $m_{c2}$ | $0.05m_0$ | [4,5] |
| ρ | 5.614 grcm$^{-3}$ | [2,5] |
| $K_s$ | $(15.0)K_0$ | [2] |
| $K_\infty$ | $(13.8)K_0$ | [2] |
| $K_0$ | $8.85\times10^{-12}C^2/Nm^2$ | [2] |
| $E_{AC}$ | 4 eV | [2] |
| $E_{NPO}$ | 6 eV | [2] |
| θ | 345 K | [2,3] |
| $N_D$ | $2\times10^{15}$ cm$^{-3}$ | [2] |
| $m_{d1}$ | $0.05m_0$ | [2] |
| $m_{d2}$ | $0.3m_0$ | [2] |
| $m_0$ | $9.11\times10^{-28}$ gr | [3] |

## 4. Results And Discussions

Using Matthiessen's rule we have calculated the theoretical hole mobility as a function of concentration at 300 K. This curve is compared with the experimental data as are shown in Fig.1. Based on the theoretical curve it seems that the mobility decreases with increasing the concentration. The agreement between the theoretical mobility and experimental results is not good. In table.2, the concentrations of the samples, especially, number 067, 613b, 617, 620, and 0826 are about $10^{17}$ – $10^{18}$ $cm^{-3}$, but their mobilities are about 91.02–230.71 $cm^2$/Vs.The result of the theoretical analysis shows that at room temperature

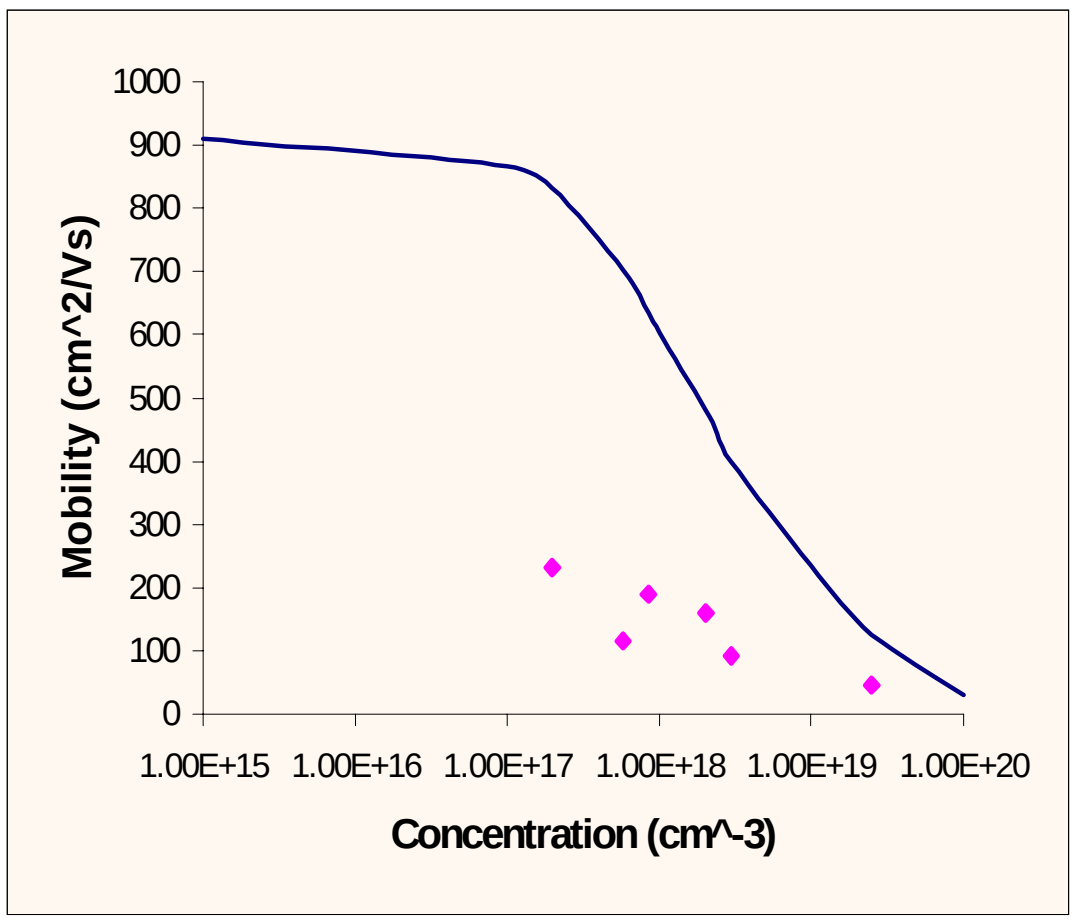

FIG.1.The hole mobility of undoped p-GaSb as a function of hole concentration with $N_D$ = $10^{15}$ $cm^{-3}$ at 300 K.(♦are the experimental results and solid curve is the theoretical analysis)

the concentration ≈ $10^{16}$ $cm^{-3}$ the mobility is about 910 $cm^2$/Vs, the concentration ≈ $10^{17}$ $cm^{-3}$ the mobility is about 850 $cm^2$/Vs, and the concentration ≈ $10^{18}$ $cm^{-3}$ the mobility is about 500 $cm^2$/Vs.

Table.2.Electrical properties of p-GaSb at 300 K obtained by Hall measurement.

| Samples | μ(hole mobility) ($cm^2$/Vs) | Hole concentration ($cm^{-3}$) |
|---|---|---|
| 67a | 230.71 | 1.99x$10^{17}$ |
| 613b | 159.81 | 2.01x$10^{18}$ |
| 617 | 188.65 | 8.52x$10^{17}$ |
| 620 | 91.02 | 3x$10^{18}$ |
| 0821 | 45.33 | 2.5x$10^{19}$ |
| 0826 | 115.2 | 5.76x$10^{17}$ |

To explain figure.1 in detail, we also calculated the individual mobility components due to the various scattering processes and the effective (total) mobility due to all of the scattering mechanisms vary with the temperature as are shown in Fig.2. The effective mobility is calculated according to Matthiessen's rule and this mobility will be compared with experimental mobility. Unfortunately, we have no the cryostat apparatus to measure the mobility at all temperatures. It can be seen that the effects of acoustic, polar optical and non-polar optical scattering are most pronounced at about room temperature. The three scattering processes result from lattice vibrations (phonon scattering). Since lattice vibration increases with increasing temperature, phonon scattering becomes dominant at high temperatures; hence the mobility decreases with increasing temperature. Contrarily, ionized impurity scattering becomes less significant at higher temperatures, even though the experimental hole concentrations are equal to the

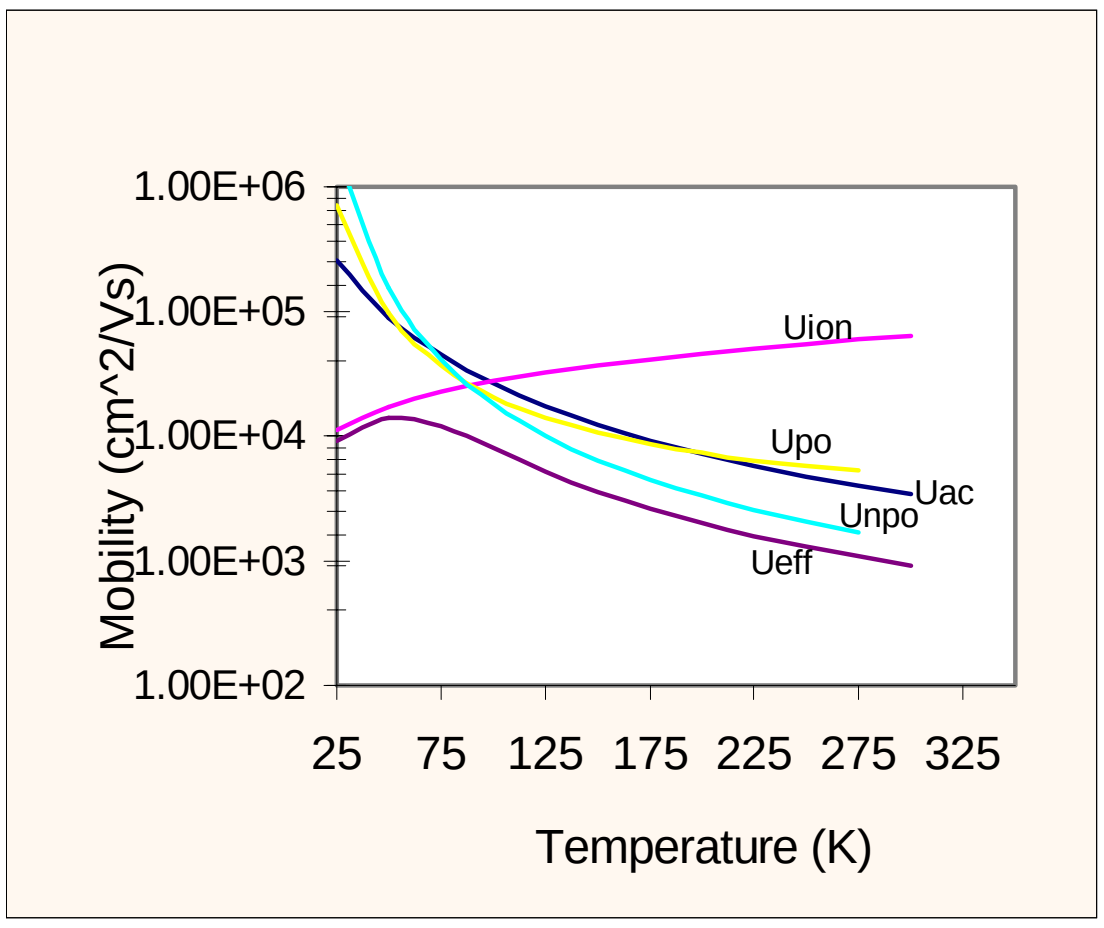

FIG.2.Temperature variation of theoretically calculated partial mobilities and effective mobility for undoped p-GaSb.(Uion, Upo, Unpo, Uac, and Ueff represent ionized impurity, polar optical, non-polar optical, acoustic, and effective mobility, respectively).

ionized scatter concentrations. At higher temperatures, the holes move faster for a shorter time and they are therefore less effectively scattered. It also seems that the effective mobility is about 905 $cm^2$/Vs with the concentration ≈ $10^{16}$ $cm^{-3}$ at room temperature. When this result is compared with the theoretical result of Fig.1, it can be seen that both the results are proportional. However, the experimental results are not good (see Fig.1) there fore we assume that the concentration of the unintentional donors $N_D$ >$10^{15}$ $cm^{-3}$ dominated the hole mobility of the thin films. The assumption we use, because to find the theoretical

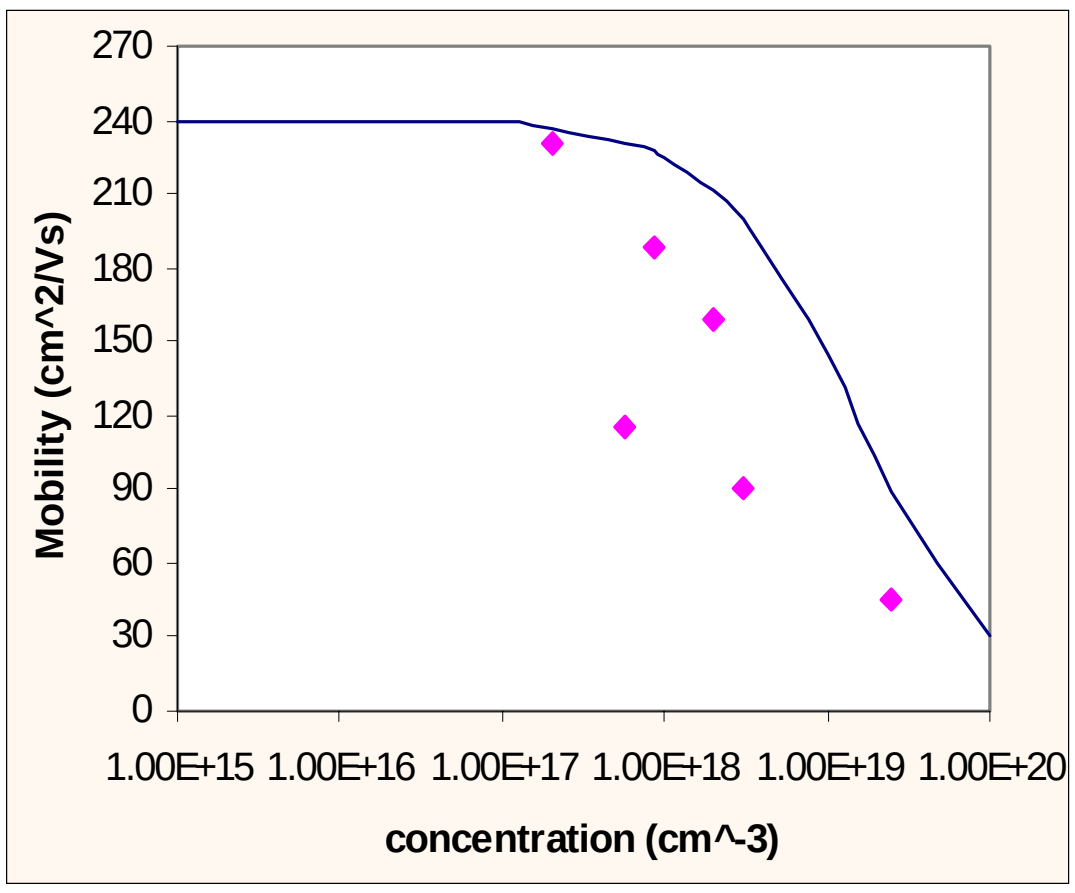

FIG.3. The hole mobility of undoped p-GaSb as a function of concentration with $N_D$ = 5.5x$10^{18}$ $cm^{-3}$ at room temperature. (♦ are the experimental results and solid curve is the theoretical analysis)

curve of figure.1 the value of $N_D$ in calculation is $10^{15}$ $cm^{-3}$. To prove this assumption, we calculated again the other theoretical curve with increasing $N_D$ = 5.5x$10^{18}$ $cm^{-3}$. This curve is then compared with experimental data as is shown in Fig.3.

According to figure.3, it seems that the comparison between experimental data and theoretical curve are suitable. This high donor concentration showing the films of p-GaSb that grown have higher native defects.

## 5. Conclusion

Theoretical analysis shows that at 300 K the mobility of undoped p-GaSb ≈ 910 $cm^2$/Vs. At low temperatures, mobility just affected by ionized impurity scattering. In contrast, at high temperatures (≈ 300 K) acoustic, polar optical, and non-polar optical scattering processes dominate the mobility. The comparison between the experimental data and theoretical curve as a function of concentration shows that the experimental mobilities are still lower. The thin films of undoped p-GaSb that grown probably have the unintentional donor concentrations are about $10^{18}$ – $10^{19}$ $cm^{-3}$. Based on the high unintentional donor concentrations that the hole mobility of undoped p-GaSb at room temperature is not only controlled by the three scattering (phonon scattering) mechanisms according to the theoretical analysis but also by the four scattering processes.

## Acknowledgements

This work was supported by Laboratory of Electronic Material Physics, Bandung Institute of Technology, Jl. Ganesa 10 Bandung, Indonesia 40132.